\documentclass[twocolumn,aps,prl,groupedaddress, nobalancelastpage]{revtex4}
\usepackage{graphicx}
\usepackage{color}
\usepackage{amsmath}
\usepackage{marvosym}
\usepackage{bm}
\usepackage{verbatim}

\usepackage{amssymb}
\usepackage{amsthm}
\usepackage{amsfonts}
\usepackage{bbm}
\usepackage{hyperref}
\usepackage{stmaryrd}

\newcommand{\be}{\begin{equation}}
\newcommand{\ee}{\end{equation}}
\newcommand{\bea}{\begin{eqnarray}}
\newcommand{\eea}{\end{eqnarray}}

\newcommand{\la}{\langle}
\newcommand{\ra}{\rangle}

\renewcommand{\vec}[1]{{\bf #1}}

\begin{document}

\title{Strain-induced large injection current in twisted bilayer graphene}
\author{Arpit Arora$^1$}
\author{Jian Feng Kong$^2$}
\author{Justin C. W. Song$^{1}$}
\email{justinsong@ntu.edu.sg}
\affiliation{$^1$Division of Physics and Applied Physics, Nanyang Technological University, Singapore 637371}
\affiliation{$^2$Institute of High Performance Computing, Agency for Science, Technology, and Research, Singapore 138632}

\begin{abstract}
The electronic wavefunctions in moir\'e materials are highly sensitive to the details of the local atomic configuration enabling Bloch band geometry and topology to be controlled by stacking and strain. Here we predict that large injection currents (under circular polarized irradiation) can develop in strained twisted bilayer graphene (TBG) heterostructures with broken sublattice symmetry; such bulk photovoltaic currents flow even in the absence of a p-n junction and can be controlled by the helicity of incident light. As we argue, large injection current rates proceed from strong and highly peaked interband Berry curvature dipole distributions (arising from the texturing of Bloch wavefunctions in strained TBG heterostructures). Strikingly, we find that TBG injection current displays pronounced responses in the THz regime and can be tuned by chemical potential. These render injection currents a useful photocurrent probe of symmetry breaking in TBG heterostructures and make TBG a promising material for THz technology.
\end{abstract} 

\maketitle

Moir\'e materials provide means to reconstruct bandstructure and engineer its correlations~\cite{geim,EQMM,andrea}. Perhaps the most prominent example is flat energy bands that arise when bilayer graphene is twisted to magic angle~\cite{neto,bm} hosting superconductivity and strongly correlated insulators~\cite{pablo1,pablo2,dean1}. Beyond the redesign of energy dispersion, however, is the ability of moir\'e materials to alter the texture of Bloch wavefunctions. For instance, twisting atomic layers allows the design of topological Bloch bands formed out of topologically trivial constituent layers~\cite{song,yao,macdonald,senthil,gordon1,young1}, as well as the control of strong concentrations of Bloch band quantum geometric quantities such as Berry curvature density~\cite{law1,pako}. Given the intimate relationship between band geometry and optical nonlinearities \cite{Morimoto}, moir\'e materials are expected to host pronounced nonlinear photocurrent responses~\cite{dai2, BHYan}.

Here we argue that twisted bilayer graphene (TBG) heterostructures can host large injection currents under irradiation of circularly polarized light that change direction with chirality of light. When TBG is strained and encapsulated with hexagonal Boron Nitride (hBN) it naturally breaks $C_{3}$~\cite{liangfustrain,law1,pako} and $C_{2}$ \cite{senthil,law1,pako2} symmetries, and as we argue below, enable injection currents to be induced. Indeed, in real TBG samples, heterostrain of 0.1 - 0.5\% have been measured in STM experiments~\cite{pasupathy,yazdani,choi,bediako}. Similarly, hBN induced $C_{2}$ symmetry breaking \cite{hunt,adam,giovannetti,hone} can lead to charge neutrality (CN) gaps of order several to tens of meV for graphene/hBN heterostructures. As a result, for moderate strain and gap values we find large injection current rates even for modest light intensities [see discussion below].

\begin{figure}[h!]
\includegraphics[width=\columnwidth]{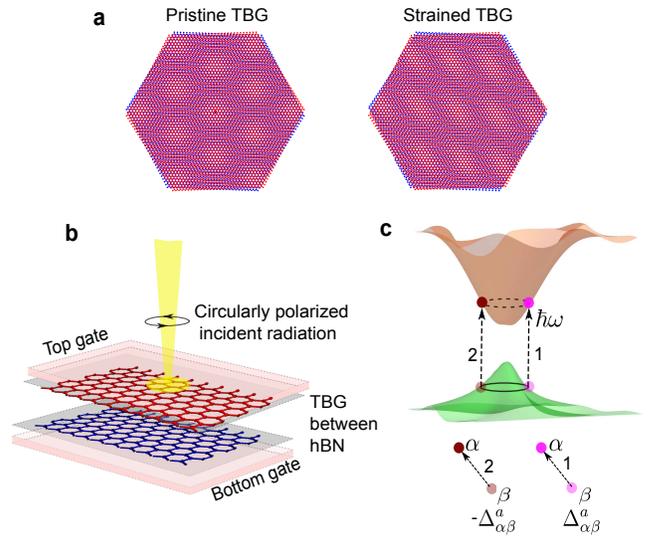}
\caption{(a) Illustration of moir\'e superlattice for pristine TBG (left panel) and strained TBG (right panel); in the latter, an elliptical moir\'e pattern is induced in TBG due to geometric deformations under strain. (b) Device geometry of strained TBG encapsulated with hBN that can display a circularly polarized photocurrent response. (c) Schematic depicting the processes that produce a (circularly polarized) injection current in TBG. Injection current arises from changes to electron velocity, $\boldsymbol{\Delta}_{\alpha\beta}(\vec{k})$ (see text), as it undergoes a photo-induced interband transition. The solid (dashed) circles indicate the initial $\beta$ (final $\alpha$) isoenergy contours that satisfy energy conservation $\delta(\hbar\omega_{\alpha\beta}(\vec{k}) - \hbar\omega)$. (bottom) Even as time reversal symmetry in TBG ensures $\Delta_{\alpha\beta}^a(\vec{k}) = -\Delta_{\alpha\beta}^a(-\vec{k})$, when inversion symmetry is broken, transition rates of process ``1'' [rate $W_{\alpha \leftarrow \beta} (\vec k)$] and ``2'' [rate $W_{\alpha \leftarrow \beta} (-\vec k)$] can be imbalanced under circular polarized irradiation. }
\label{fig1}
\end{figure}

Strikingly, we find TBG injection currents are dominated by transitions between flat and remote bands. This can be traced to strong interband Berry curvature dipole (IBCD) densities for these transitions and its strain induced asymmetric peak-like distribution across the moir\'e Brillioun zone (mBZ). This highlights the dramatic effect of Bloch wavefunction texture in TBG heterostructures. Interestingly, the distribution of IBCD also leads to an injection current that is highly sensitive to chemical potential allowing injection current to be gate-controlled in TBG. Given the sensitivity of the formation of correlated states in TBG to the presence of (extrinsically) broken symmetries (e.g., ferromagnetism can be found in hBN encapsulated TBG~\cite{gordon1,young1}), we anticipate that the large injection currents can be used as sensitive photocurrent probe of broken $C_{2}$ and $C_{3}$ symmetries in TBG systems.

\textit{Injection current and symmetry}: Injection current is a class of bulk photovoltaic effect that arises from the inversion asymmetric changes to carrier velocity during an interband optical excitation \cite{sipe,nagaosa}, see Fig.~\ref{fig1}b,c. It is described as a DC second order response in applied electric field and, therefore, only develops in inversion asymmetric materials. For an electronic charge $-e(e>0)$, the injection current rate $\partial_t \vec{j}$ \cite{nagaosa} can be obtained by tracking the carrier's change in velocity $\vec{\Delta}_{\alpha\leftarrow\beta}$ during interband transition from an initial state $\beta$ to final state $\alpha$ [Fig.~\ref{fig1}c], i.e., 
\begin{equation}
\label{eq:transitions}
\partial_t j^a = -e\sum_{\alpha\leftarrow\beta}W_{\alpha\leftarrow\beta} \Delta_{\alpha\leftarrow\beta}^a, 
\end{equation}
where Latin and Greek indices denote directions and bands respectively, and $W_{\alpha\leftarrow\beta}$ is the transition rate.
 
The interband transition rate for an electronic system with Bloch Hamiltonian $H(\vec{k})|u_\alpha(\vec{k})\ra = \hbar\omega_{\alpha}(\vec{k})|u_\alpha(\vec{k})\ra$, can be described by Fermi's golden rule $W_{\alpha\leftarrow\beta} = [2\pi/\hbar]f_\beta[1 - f_\alpha]|V_{\alpha\leftarrow\beta}|^2 \delta(\hbar\omega_{\alpha\beta} - \hbar\omega)$, where $f_{\alpha(\beta)} = f(\hbar\omega_{\alpha(\beta)})$ is the Fermi function and $\omega_{\alpha\beta} = \omega_\alpha - \omega_\beta$. Writing the electric field $\vec{E}(t) = \vec{E} e^{-i\omega t} + c.c$, the matrix element reads $V_{\alpha\leftarrow\beta} = \sum_b eE_b r_{\alpha\beta}^b$, where $r_{\alpha\beta}^b$ is the interband Berry connection. We note that $r_{\alpha\beta}^b = v_{\alpha\beta}^b/(i\omega_{\alpha\beta})$ for $\alpha \neq \beta$; here $v_{\alpha\beta}^b=\hbar^{-1}\la u_\alpha|\partial_{k_b}H|u_\beta\ra$ is the velocity matrix element. As a result, the injection current rate for circularly polarized irradiation [with $\vec E = E_x \hat{\vec x} - i E_y \hat{\vec y}$] is \cite{nagaosa,sipebook,grushin}
\begin{align}
\partial_t j^a  & = 
i C \sum_{\vec k, \alpha \neq \beta} f_{\alpha\beta}\Delta^a_{\alpha\beta} \llbracket r^x_{\alpha\beta}, r^y_{\beta \alpha}\rrbracket\delta(\omega_{\alpha\beta}-\omega) E_x E_y \label{eq:injectioncurrent}\\
& = - C \sum_{\vec k, \alpha \neq \beta} \mathcal{M}^{axy}_{\alpha \beta} (\vec k) f_{\alpha\beta} \delta(\omega_{\alpha\beta}-\omega) E_x E_y\label{eq:ibcd}
\end{align}
where $C = 2\pi e^3/\hbar^2$, the $k$ space integral reads $\sum_{\vec k} 
= \int d^dk/(2\pi)^d$, $\llbracket r^{x}_{\alpha\beta}, r^{y}_{\beta \alpha}\rrbracket = r^{x}_{\alpha\beta} r^{y}_{\beta \alpha} - r^{x}_{\beta\alpha} r^{y}_{\alpha\beta}$, the velocity difference is $\Delta^a_{\alpha\beta} = v_{\alpha\alpha}^a - v_{\beta\beta}^a$, and $f_{\alpha\beta} = f_\alpha - f_\beta$. In the second line, we have written $\mathcal{M}^{axy}_{\alpha\beta} (\vec k)= - i  \llbracket r^{x}_{\alpha\beta}, r^{y}_{\beta \alpha}\rrbracket \Delta^a_{\alpha\beta} $ as the {\it interband} Berry curvature dipole (IBCD) to highlight its role in determining injection currents, see below. From a physical point of view, IBCD captures the strength of the injection current (rate) for each transition $\beta, \vec k \to \alpha, \vec k$. We note, parenthetically, that injection currents are non-zero only for circularly polarized light \cite{nagaosa,Morimoto} in the presence of time reversal symmetry.

Nonlinear photocurrents are particularly sensitive to crystalline symmetry. For example, in the presence of inversion symmetry, $\hat{\mathcal{I}} H(\vec{k}) \hat{\mathcal{I}}^{-1} = H(-\vec{k})$ yielding $\omega_\alpha(-\vec{k}) = \omega_\alpha(\vec{k}) $ and $v_{\alpha\beta}^b(-\vec{k}) = -v_{\alpha\beta}^b(\vec{k})$ [see Fig. \ref{fig1}c]. As a result, Eq.(\ref{eq:transitions}) and Eq.~(\ref{eq:injectioncurrent}) vanish under inversion symmetry. While pristine TBG possesses inversion symmetry, hBN aligned encapsulated TBG breaks inversion symmetry by lifting $C_{2}$ symmetry (AB sublattice symmetry) to produce a gap at charge neutrality (CN).

Even when $C_{2}$ symmetry is broken, thereby enabling a Berry curvature distribution to develop, an in-plane [$x$-$y$ plane] injection current arising due to normal incident ($z$-direction) irradiation is still forbidden by $C_{3}$ symmetry. This can be seen by analysing Eq. (\ref{eq:injectioncurrent}) at wave vectors related by the three-fold rotation matrix 
\begin{equation}
\mathcal{R} = \begin{pmatrix}
\cos\frac{2\pi m}{3} & -\sin\frac{2\pi m}{3} \\
\sin\frac{2\pi m}{3} & \cos\frac{2\pi m}{3}
\end{pmatrix}; \quad m\in \{0,1,2\}
\end{equation}
satisfying $k_a = \mathcal{R}_{ja}k_{j}$. In the presence of rotational symmetry the Bloch hamiltonian satisfies $\hat{\mathcal{R}}H(\vec{k})\hat{\mathcal{R}}^{-1} = H(\mathcal{R}\vec{k})$, so that $\hat{\mathcal{R}}|u_\alpha(\vec{k})\ra = |u_\alpha(\mathcal{R}\vec{k})\ra$, with energies obeying $\omega_\alpha(\vec{k}) = \omega_\alpha(\mathcal{R}\vec{k})$ and $v_{\alpha\beta}^b (\mathcal{R}\vec{k}) = \mathcal{R}_{bj}v_{\alpha\beta}^{j} (\vec{k})$ \cite{nagaosa}. Given these symmetry relations, we find that the factors in Eq.~(\ref{eq:injectioncurrent}) are related via $\llbracket r^b_{\alpha\beta}(\vec{k}),r^c_{\beta\alpha}(\vec{k})\rrbracket = \llbracket r^b_{\alpha\beta}(\mathcal{R}\vec{k}),r^c_{\beta\alpha}(\mathcal{R}\vec{k}) \rrbracket$ and $\Delta_{\alpha\beta}^a (\mathcal{R}\vec{k}) = \mathcal{R}_{aj}\Delta_{\alpha\beta}^{j} (\vec{k})$. Summing across triplets of wavevectors $\vec k$ related by $\mathcal{R}$ yields a vanishing injection current in the presence of $C_{3}$ symmetry. However, broken $C_{3}$ symmetry can be naturally achieved in realistic TBG samples via strain enabling a finite injection current to manifest. 

\begin{figure*}
\centerline{\includegraphics[width=\textwidth]{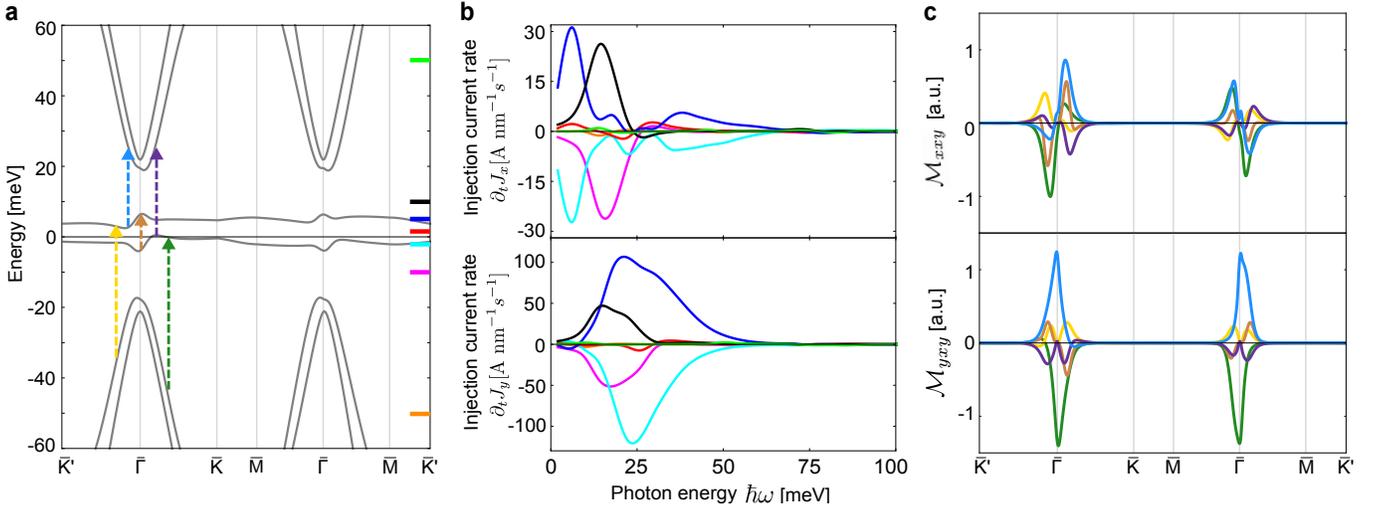}}
\caption{(a) Bandstructure of strained TBG with hBN encapsulation focussing on the bands close to charge neutrality at twist angle $\theta=1.05^\circ$. (b) Injection current rate under circular polarized irradiation as a function of incident photon energy plotted at different chemical potentials. Colored curves in panel (b) denote $\mu$ [in meV] $= -50$ (orange), $-10$ (magenta), $-2$ (cyan), $1.5$ (red), $5$ (blue), $10$ (black) and $50$ (light green) [these chemical potential levels are also indicated in panel (a) with bars of the same color on the right axis]. (c) Interband Berry curvature dipole (IBCD) distributions $\mathcal{M}_{\alpha\beta}^{abc} (\vec k)$ corresponding to vertical transitions $\beta \to \alpha$ indicated by colored arrows in panel (a). Parameters used: heterostrain $\epsilon = 0.1\%$, with a staggered sublattice potential of $\Delta_1 = \Delta_2 = 5$ meV, and electric field amplitude, $\vec{E} = E_0(1,-i)$ with $E_0 = 100$ V/cm.
}
\label{fig2}
\end{figure*}

\textit{Continuum model for strained TBG-hBN heterostructures}: The low energy physics for small twist angle TBG can be described by a continuum model formed by massless Dirac fermions in each layer \cite{neto,bm,koshinoPRX}. The TBG Hamiltonian for valley $\zeta = \pm$ is 
\begin{equation}
\label{eq:tbghamiltonian}
H_\zeta = \begin{pmatrix}
H_{1,\zeta}(\vec{q}) & U_\zeta^\dag \\
U_\zeta & H_{2,\zeta}(\vec{q})
\end{pmatrix}
\end{equation}
where $l=1,2$ is the layer index for the two graphene layers rotated by $\mp\theta/2$ about the normal. The Hamiltonian for each layer is 
\begin{equation}
\label{eq:monolayerhamiltonian}
H_{l,\zeta}(\vec{q}) = -\hbar v_F R(\pm\theta/2)\vec{q}.(\zeta \sigma_x,\sigma_y)
\end{equation}
where $v_F$ is the original Fermi velocity so that $\hbar v_F/a = 2135.4$ meV ($a$=0.246 nm is graphene lattice constant), $R(\theta)$ is a rotation matrix and $\sigma_\mu$ is the Pauli matrix acting on sublattice space. The wavevector $\vec{q} = \vec{k}-\vec{K}_{l,\zeta}$ is taken with respect to the original BZ Dirac point, $\vec{K}_{l,\zeta}$. The interlayer coupling between the two graphene layers reads~\cite{bm,koshinoPRX}
\begin{multline}
\label{eq:interlayer}
U_\zeta = \begin{pmatrix}
u & u' \\
u' & u 
\end{pmatrix} +
\begin{pmatrix}
u & u'e^{-i\zeta\frac{2\pi}{3}} \\
u'e^{i\zeta\frac{2\pi}{3}} & u 
\end{pmatrix}e^{i\zeta \vec{g}_1.\vec{r}} \\
+ 
\begin{pmatrix}
u & u'e^{i\zeta\frac{2\pi}{3}} \\
u'e^{-i\zeta\frac{2\pi}{3}} & u 
\end{pmatrix}e^{i\zeta (\vec{g}_1+\vec{g}_2).\vec{r}}
\end{multline}
where $\vec{g}_i$ is the reciprocal lattice vector of mBZ. In what follows, we use the tunnelling parameters $u=79.7$ meV and $u'=97.5$ meV to account for lattice relaxation in our calculations~\cite{koshinoPRX}. When hBN is aligned with the graphene layers, $C_2$ symmetry is broken modifying the layer Hamiltonians $H_{l,\zeta}$. This can be described by introducing a sublattice staggered potential $\Delta_l$ so that the Hamiltonian for each layer $H_{l,\zeta}(\vec{q}) \rightarrow H_{l,\zeta}(\vec{q})+\Delta_l\sigma_z$ \cite{pako2}.

Finally, the presence of a uniaxial heterostrain in TBG of magnitude $\epsilon$ can be described by the linear strain tensor~\cite{liangfustrain}
\begin{equation}
\label{eq:strain}
\mathcal{E}_l = \mathcal{F}(l)\epsilon\begin{pmatrix}
-\cos^2\varphi + \nu \sin^2\varphi & (1+ \nu) \cos\varphi\sin\varphi \\
(1+ \nu) \cos\varphi\sin\varphi & \nu\cos^2\varphi - \sin^2\varphi
\end{pmatrix}
\end{equation}
where $\mathcal{F}(l=1,2)=\mp1/2$, $\nu=0.165$ is the Poisson ratio of graphene and $\varphi$ gives direction of the applied strain. The strain tensor satisfies general transformations in each layer, $\vec{a}_l \rightarrow \vec{a}_l'= [\mathbbm{1}+\mathcal{E}_l]\vec{a}_l$ and $\vec{b}_l \rightarrow \vec{b}_l ' \approx [\mathbbm{1}-\mathcal{E}_l^T]\vec{b}_l$ for real and reciprocal lattice vectors respectively \cite{liangfustrain}. The strain induced geometric deformations affect the interlayer coupling [see Fig.~\ref{fig1}a] and further changes the electron motion via gauge field $\vec{A}_l = \sqrt{3}\beta/2a(\mathcal{E}^{xx}_l+\mathcal{E}^{yy}_l, -2\mathcal{E}^{xy}_l )$, where $\beta = 3.14$. As a result, we have $\vec{q} \rightarrow \vec{q}_{l,\zeta} = [\mathbbm{1}+\mathcal{E}_l^T][\vec{k}-\mathcal{K}_{l,\zeta}]$ with $\mathcal{K}_{l,\zeta} = [\mathbbm{1}-\mathcal{E}_l^T]\vec{K}_{l,\zeta} - \zeta \vec{A}_l$ \cite{liangfustrain}. 

The effective TBG Hamiltonian in Eq. (\ref{eq:tbghamiltonian}), modified by the effects of strain and hBN alignment with graphene layers via sublattice staggered potential, can be re-written as 
\begin{equation}
\label{eq:tbghamiltonianstrainandhbn}
\mathcal{H}_\zeta = \begin{pmatrix}
H_{1,\zeta}(\vec{q}_{1,\zeta})+\Delta_1\sigma_z & U_\zeta^\dag \\
U_\zeta & H_{2,\zeta}(\vec{q}_{2,\zeta})+\Delta_2\sigma_z
\end{pmatrix}
\end{equation}
Note that for a given $\vec{q}$ in the mBZ, the 4$\times$4 Hamiltonian in Eq. (\ref{eq:tbghamiltonianstrainandhbn}) is cast into a multiband eigensystem problem as the interlayer coupling leads to hybridisation of the eigenstates at Bloch vectors $\vec{q}$ and $\vec{q}'=\vec{q}+\vec{g}$, where $\vec{g} = m_1\vec{g}_1 +m_2\vec{g}_2$ and $m_{1,2}\in\mathbb{Z}$ \cite{koshinoPRX}. We truncate the size of the matrix by defining a circular cut-off $|\vec{q}-\vec{q}'|<4|\vec{g}_1|$ \cite{koshinoPRX}. The band structure for TBG ($\theta = 1.05^\circ$) at moderate strain ($\epsilon = 0.1\%$) and sublattice staggered potential $\Delta_1 = \Delta_2 = 5$ meV is shown in Fig.~\ref{fig2}a.

\textit{Large Injection current in TBG}: We numerically calculate the rate of injection current for a circularly polarized normal incident radiation $\vec{E} = E_0(1,-i)$ on TBG by using Eq. (\ref{eq:injectioncurrent}) and summing across both valleys. As an illustration of TBG injection current at small twist angles we have used $\theta=1.05^\circ$ and taken $\rm{T}=0$ K. In Fig.~\ref{fig2}b, we plot the injection current rate and its dependence on incident photon energy across different chemical potentials ranging from within the flat bands to far inside the remote valence (below CN) and remote conduction (above CN) bands. 

Strikingly, we find pronounced peaks of injection current rate with large magnitudes at photon energy in the THz regime [see Fig.~\ref{fig2}b]. Indeed peak injection current susceptibilities $\tilde{\chi}_{abc}$ (where $\partial_t j_a = \tilde{\chi}_{abc}[\vec E]_b [\vec E]^*_c$ \cite{sipe,sciadv}) can reach very large values of $\tilde{\chi}_{yxy}\approx10^{12}$ A nm V$^{-2}$s$^{-1}$; these values are two orders of magnitude larger than that reported recently for the 2D ferroelectric GeS~\cite{sciadv}. While large injection current rates can be achieved at low frequencies, we find $\partial_t \vec j$ rapidly diminishes at larger photon energy ($\sim 100 \, {\rm meV}$). 

To understand the photon energy dependence in Fig.~\ref{fig2}b we first note that the injection current rate critically depends on the velocity difference $\Delta_{\alpha\beta}^a$ (accrued by the electron in the optical transition from $\beta \to \alpha$) as well as the band resolved Berry curvature $-i\llbracket r_{\alpha\beta}^{x},r_{\beta\alpha}^y\rrbracket$. Their product yields IBCD, see Eq.~(\ref{eq:ibcd}). IBCD is the {\it interband} analog of the conventional Berry curvature dipole density (within a single band) more commonly known in the context of the nonlinear Hall effect~\cite{bcdweyl,inti}; recently, nonlinear Hall effects that arise from large berry curvature dipoles within a single band have been found in strained TBG heterostructures~\cite{pako,law1}.  Unlike that found in the nonlinear Hall effect, however, injection current rate is most sensitive to regions of IBCD corresponding to the interband transition isoenergy contours [defined by $\delta(\omega_{\alpha\beta} (\vec k)- \omega)$ in Eq.~(\ref{eq:injectioncurrent})]. As a result, its distribution provides critical information on the frequency dependence of TBG injection current.

This dependence is exemplified in Fig.~\ref{fig2}c where we plot the IBCD [$\mathcal{M}^{(x,y)xy}_{\alpha\beta}$] densities for representative $\beta \to \alpha$ transitions between flat $\to$ flat and flat/remote $\to$ remote/flat bands; here the IBCD pattern correspond to interband transitions denoted by the (colored) arrows in Fig.~\ref{fig2}a. Interestingly, IBCD peaks close to the $\bar{\Gamma}$ point, rapidly diminishing as $\vec k$ moves away from $\bar{\Gamma}$. Inspecting the TBG bandstructure in Fig.~\ref{fig2}a, we find these transitions correspond to low photon energies of order tens of meV consistent with the large injection current peaks in the THz regime, see Fig.~\ref{fig2}b. Indeed, this indicates how TBG injection current is dominated by the large IBCD found for transitions between flat bands and remote bands. In contrast, when photon energies are large, TBG injection current rates are dramatically suppressed (see Fig.~\ref{fig2}b). This is consistent with the small IBCD distributions for such transition energies, see e.g., Fig.~\ref{fig2}c. 

The complex pattern of IBCD (as a function of $\vec k$ as well as for different interband $\beta \to \alpha$ transitions) found in Fig.~\ref{fig2}c suggests that $\partial_t\vec j$ can be highly sensitive to chemical potential, see curves in Fig.~\ref{fig2}b displaying injection current rate for various chemical potential values. For instance, $\partial_t \vec j$ displays a very weak response when chemical potential is tuned close to CN [see red injection current rate, Fig.~\ref{fig2}b, where we have used $\mu = 1.5 \, {\rm meV}$ inside the gap between the flat conduction (FC) and flat valence (FV) bands]. The suppressed response arises despite the presence of accessible low energy and large values of $\mathcal{M}^{(x,y)xy}_{\alpha\beta}$ close to $\bar{\Gamma}$. This unusual situation can be understood from the IBCD distributions for the various transitions e.g., remote valence (RV) bands $\to$ flat bands, flatbands $\to$ remote conduction band (RC) and FV $\to$ FC bands, see IBCD distributions in Fig. 2c. When accounting for these contributions in Eq.~(\ref{eq:ibcd}) we obtain a suppressed response. 

However, when chemical potential is tuned slightly away from CN, partial compensation is avoided and large injection current rates are turned ``on'', Fig.~\ref{fig2}b. For example, when chemical potential is fixed inside the gap between FC and RC bands (see $\mu =10 \, {\rm meV}$, Fig.~\ref{fig2}a,b), large injection currents manifest. These currents are dominated by transitions from the FC to RC bands (blue arrow) that corresponds to the blue IBCD distribution in Fig.~\ref{fig2}c.

\begin{figure}
\includegraphics[scale=0.072]{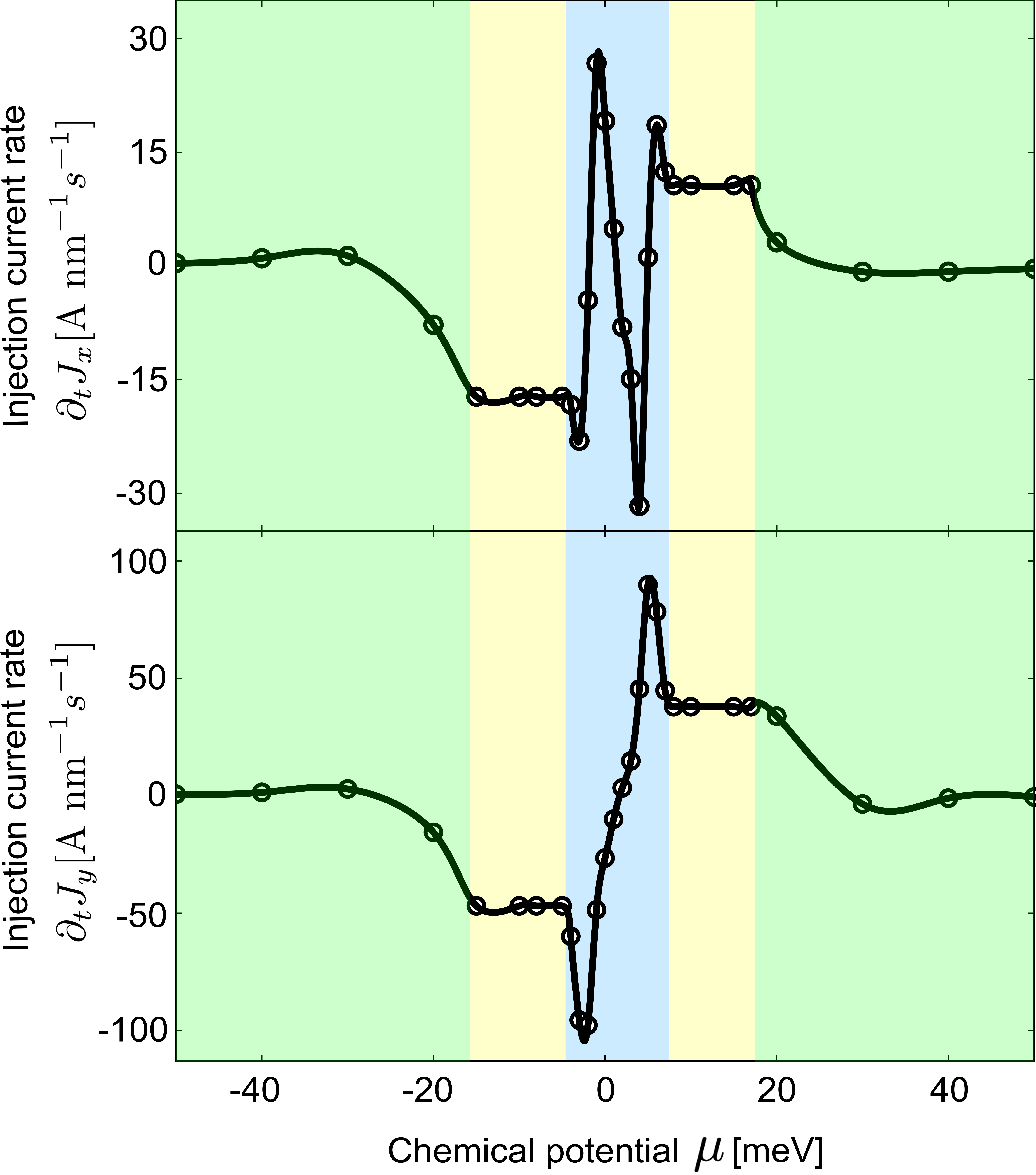}
\caption{Variation of injection current rate (under circular polarized irradiation) in strained and hBN encapsulated TBG as a function of chemical potential; green, yellow and, blue regions denote regions where the chemical potential is in the remote bands, in the gap between remote and flat bands, and within the flat band region respectively. Here we have fixed the incident photon energy $\hbar\omega=20$ meV as an illustration. Parameters used are the same as Fig.~\ref{fig2}.}
\label{fig3}
\end{figure}

In a similar fashion, when chemical potential is fixed between the RV and FV bands (see $\mu =-10 \, {\rm meV}$, Fig.~\ref{fig2}a,b), injection current is dominated by transitions from the RV to FV bands (dark green arrow) that corresponds to the dark green IBCD distribution in Fig.~\ref{fig2}c. Strikingly, injection current for $\mu =-10 \, {\rm meV}$ has an opposite sign to that of $\mu =10 \, {\rm meV}$ (see Fig.~\ref{fig2}b) arising from the opposite signs of the integrated IBCD corresponding to their constituent transitions (blue vs dark green IBCD distributions in Fig.~\ref{fig2}c). A more detailed investigation of injection current variation with chemical potential is shown in Fig.~\ref{fig3} (here we have fixed photon energy at $\hbar \omega = 20\, {\rm meV}$ as an illustration). In the same fashion as above, this dependence can be understood from the IBCD distribution in Fig.~\ref{fig2}c; it vividly displays how the position of the chemical potential can enable control over the allowable interband transitions to produce a gate-tunable TBG injection current.

In summary, we find large circular injection currents can be produced in hBN encapsulated-TBG under modest strain values. Arising from the large IBCD for transitions between flat and remote bands, injection currents display especially pronounced response in the THz regime for small twist angle TBG, as discussed here. This makes it an interesting candidate material for gate-tunable THz photodetection and circuits \cite{BHYan}. Given the strong absorption characteristics of TBG moire materials \cite{koshinoPRB,dai2} (e.g., those corresponding to transitions involving the flat bands), we anticipate that large TBG injection currents can proliferate across a multitude of twist angles.

{\it Acknowledgments}: We thank Frank Koppens and Roshan Krishna Kumar for useful conversations. This work was supported by Singapore MOE Academic Research Fund Tier 3 Grant MOE2018-T3-1-002 and a Nanyang Technological University start-up grant (NTU-SUG).

\newpage
\onecolumngrid
\renewcommand{\theequation}{S\arabic{equation}}
\renewcommand{\thefigure}{S\arabic{figure}}
\renewcommand{\thetable}{S\Roman{table}}
\makeatletter
\makeatother
\setcounter{equation}{0}
\setcounter{figure}{0}

\begin{center}
\textbf{Supplementary Information for\\
 ``Strain-induced large injection current in twisted bilayer graphene"}
\end{center}
\twocolumngrid
\subsection{Encapsulated TBG with Heterostrain Strain}

In the main text, we used a continuum model to describe the bandstructure of twisted bilayer graphene (TBG)~ \cite{neto,bm,koshinoPRX,koshinoPRB} under strain \cite{liangfustrain} and aligned with hBN~\cite{pako2}. For the convenience of the reader, here we review the continuum model and describe how strain and sublattice symmetry breaking is implemented in the Hamiltonian.

\subsubsection{TBG Lattice Structure and Continuum Hamiltonian}
For TBG, we define the lattice structure as in Ref. \cite{koshinoPRX}. In each graphene layer the primitive (original) lattice vectors are $\vec{a}_1 = a(1,0)$ and $\vec{a}_2 = a(1/2,\sqrt{3}/2)$ with $a=0.246$ nm being the lattice constant. The corresponding reciprocal space lattice vectors are $\vec{b}_1 = (2\pi/a)(1,-1/\sqrt{3})$ and $\vec{b}_2 = (2\pi/a)(0,2/\sqrt{3})$, and Dirac points are located at $K_\zeta = -\zeta(2\vec{b}_1 +\vec{b}_2)/3$. For a twist angle $\theta$ (accounting for the rotation of layers), the lattice vectors of layer $l$ are given by $\vec{a}_{l,i} = R(\mp\theta/2)\vec{a}_i$, $\mp$ for $l=1,2$ respectively, and $R(\theta)$ represents rotation by an angle $\theta$ about the normal. Also, from $\vec{a}_{l,i}.\vec{b}_{l',j} = 2\pi \delta_{ij}\delta_{ll'}$ we can check that the reciprocal lattice vectors become $\vec{b}_{l,i} = R(\mp\theta/2)\vec{b}_i$ with corresponding Dirac points now located at $\vec{K}_{l,\zeta} =  -\zeta(2\vec{b}_{l,1} +\vec{b}_{l,2})/3$.

At small angles, the slight mismatch of the lattice period between two layers gives rise to long range moir\'e superlattices. The reciprocal lattice vectors for these moir\'e superlattices are given as $\vec{g}_i = \vec{b}_{1,i} - \vec{b}_{2,i}$. The superlattice vectors $\vec{L}$, can then be found using $\vec{g}_i.\vec{L}_j = 2\pi \delta_{ij}$, where $\vec{L}_1$ and $\vec{L}_2$ span the moir\'e unit cell with lattice constant $L = \vec{L}_1 = \vec{L}_2 = a/[2\sin\theta/2]$. 

Next, when the moir\'e superlattice constant is much longer than the atomic scale, the electronic structure can be described using an effective continuum model for each valley $\zeta=\pm$. The total Hamiltonian is block diagonal in the valley index, and for each valley effective Hamiltonian of the continuum model is written in terms of the sublattice and layer basis $(A_1, B_1, A_2, B_2)$ \cite{koshinoPRX} [also see Eq. (\ref{eq:tbghamiltonian}) in main text]
\begin{equation}
H_\zeta = 	\begin{pmatrix}
H_{1,\zeta}(\vec{q}) & U^\dag \\
U & H_{2,\zeta}(\vec{q})
\end{pmatrix}
\end{equation}
where $H_{l,\zeta} = -\hbar v_F R(\pm \theta/2)\vec{q}.(\zeta\sigma_x, \sigma_y)$ is the Hamiltonian for each layer and 
\begin{multline}
U_\zeta = \begin{pmatrix}
u & u' \\
u' & u 
\end{pmatrix} +
\begin{pmatrix}
u & u'e^{-i\zeta\frac{2\pi}{3}} \\
u'e^{i\zeta\frac{2\pi}{3}} & u 
\end{pmatrix}e^{i\zeta \vec{g}_1.\vec{r}} \\
+ 
\begin{pmatrix}
u & u'e^{i\zeta\frac{2\pi}{3}} \\
u'e^{-i\zeta\frac{2\pi}{3}} & u 
\end{pmatrix}e^{i\zeta (\vec{g}_1+\vec{g}_2).\vec{r}}
\end{multline}
is the interlayer coupling with $u=79.7$ meV and $u'=97.5$ meV \cite{koshinoPRX}, thus accounting for lattice relaxation effects. TBG as modelled above is inversion symmetric, has Dirac band crossings protected by $C_2 T$, as well as $C_3$ symmetry~\cite{law1,pako,pako2,koshinoPRX}. Note that aligning hBN with graphene layers breaks $C_2$ symmetry and applying strain breaks $C_{3}$ symmetry in TBG, thus allowing a finite (circular) injection current as discussed in the main text. 

\subsubsection{TBG-hBN heterostructures}
When graphene is stacked with hexagonal Boron Nitride, A/B sublattice symmetry can be broken leading to sizeable gaps opening at the Dirac point that have been measured of order several to tens of meV \cite{hunt,hone}.
Similarly, when TBG is stacked with hBN, sublattice symmetry can be broken breaking inversion symmetry of TBG and opening a gap up at CN~\cite{pako2}.  In our model, we capture this by introducing a sublattice staggered potential~\cite{pako} inducing the breaking of AB sublattice symmetry in each layer and the Dirac Hamiltonian in each layer changes as 
\begin{equation}
H_{l,\zeta} (\vec{q}) \rightarrow H_{l,\zeta} (\vec{q}) + \Delta_l \sigma_z
\end{equation}
The alignment of hBN with graphene layers breaks all symmetries of pristine graphene except $C_{3}$ symmetry \cite{law1}. 
\subsubsection{TBG with Uniaxial strain}
We model heterostrain strain in TBG by following Ref. \cite{liangfustrain}. First, we consider uniaxial strain on each graphene layer so that a compressive force along, say $x$-direction describes the longitudinal contraction and transverse relaxation given by the strain tensor \cite{vitor}
\begin{equation}
\mathcal{E} = \epsilon\begin{pmatrix}
-1 & 0 \\
0 & \nu
\end{pmatrix}
\end{equation}
where $\epsilon$ is the strain magnitude and $\nu$ is the Poisson ratio (for graphene $\nu=0.165$ \cite{vitor}). For a force along any other direction, it is straightforward to generalise the strain tensor \cite{vitor}
\begin{equation}
\mathcal{E} = \epsilon\begin{pmatrix}
-\cos^2\varphi + \nu \sin^2\varphi & (1+ \nu) \cos\varphi\sin\varphi \\
(1+ \nu) \cos\varphi\sin\varphi & \nu\cos^2\varphi - \sin^2\varphi
\end{pmatrix}
\end{equation}
where $\varphi$ gives the direction of strain.
 
Now for TBG, we implement heterostrain by applying strain in opposite directions in two graphene layers, and thus writing the layer dependent  strain tensor \cite{liangfustrain}
\begin{equation}
\label{eq:strainS}
\mathcal{E}_l = \mathcal{F}(l) \mathcal{E}
\end{equation}
where $\mathcal{F}(l=1,2)=\mp1/2$. The strain tensor in Eq.~(\ref{eq:strainS}), also see Eq.~(\ref{eq:strain}) in main text, modifies the real and reciprocal lattice vectors in each layer
\begin{eqnarray}
\vec{a}_{i,l}&\rightarrow& \vec{a}_{i,l}'=(1+\mathcal{E}_l)\vec{a}_{i,l} \nonumber\\
\vec{b}_{i,l} &\rightarrow & \vec{b}_{i,l}' \approx  (1-\mathcal{E}^T_l)\vec{b}_{i,l} 
\end{eqnarray}
which leads to a geometrical distortion of the lattice. This makes the moir\'e dots appear elliptical~\cite{pasupathy,yazdani,choi,bediako}.

Secondly, in addition to geometric effects, strain can induce gauge fields~\cite{vozmediano} 
\begin{equation}
\vec{A}_l = \sqrt{3}\beta/2a(\mathcal{E}^{xx}_l+\mathcal{E}^{yy}_l, -2\mathcal{E}^{xy}_l )
\end{equation}
which captures the shift in the location of low energy Dirac fermions away from the rescaled valley points as the new Dirac points are now at 
\begin{equation}
\mathcal{K}_{l,\zeta} = [\mathbbm{1}-\mathcal{E}_l^T]\vec{K}_{l,\zeta} - \zeta \vec{A}_l.
\end{equation}
This further modifies the Hamiltonian for Dirac electrons in each layer
\begin{equation}
H_{1,\zeta}(\vec{q}_{1,\zeta}) = -\hbar v_F R(\pm \theta/2)\vec{q}_{1,\zeta}.(\zeta\sigma_x, \sigma_y)
\end{equation}
where the momentum exchange is now $\vec{q}_{l,\zeta} = [\mathbbm{1}+\mathcal{E}_l^T][\vec{k}-\mathcal{K}_{l,\zeta}]$.

Note that strain (alone) breaks all symmetries relevant to pristine TBG except $C_{2}$ symmetry and thus preserving the crossing of Dirac bands due to $C_{2}\mathcal{T}$ symmetry. However, the Dirac crossings are now located at generic points away from Brillouin zone corners because of the lack of threefold rotational symmetry in the presence of strain \cite{liangfustrain}.

\subsubsection{Final Hamiltonian}
TBG Hamiltonian accounting for alignment with hBN and heterostrain strain is [see also (\ref{eq:tbghamiltonianstrainandhbn}) in main text]
\begin{equation}
\mathcal{H}_\zeta = \begin{pmatrix}
H_{1,\zeta}(\vec{q}_{1,\zeta})+\Delta_1\sigma_z & U_\zeta^\dag \\
U_\zeta & H_{2,\zeta}(\vec{q}_{2,\zeta})+\Delta_2\sigma_z
\end{pmatrix}
\end{equation}
which breaks all point group spatial symmetries relevant to pristine TBG. As mentioned in the main text, the hybridization of eigenstates at Bloch vector $\vec{q}$ hybridise with that at $\vec{q}' = \vec{q}+\vec{g}$ ($\vec{g} = m_1 \vec{g}_1 + m_2 \vec{g}_2$; $m_{1,2}\in \mathbb{Z}$) due to the coupling of two graphene layers. We truncate the size of the matrix (for numerical evaluation) by including point within $|\vec{q}-\vec{q}'|=|\vec{g}|<4|\vec{g}_1|$ or $4|\vec{g}_2|$~\cite{koshinoPRX}. For a given Bloch vector $\vec{q}$, this gives us 61 sites in reciprocal space, and a corresponding matrix of size $244 \times 244$ which is then diagonalized to obtain eigenvalues and eigenvectors for numerical evaluation of injection current. 

\subsection{Numerical evaluation of injection current}
The integrals in Eq. (\ref{eq:injectioncurrent}) of main text to evaluate injection current are calculated as Riemann sum over discrete grid in $(k_x, k_y)$ plane of moir\'e Brillouin zone (mBZ) in each valley (contribution from the two valleys are then added). All the evaluations were carried out on a grid of $400\times 400$ points. Further, the $\delta$ function is approximated as a Lorentzian with phenomenological energy broadening of 3 meV~\cite{koshinoPRB}. This value of broadening parameter as well as Lorentizian distribution was recently used to successfully describe the optical absorption transitions in TBG~\cite{koshinoPRB}.

\end{document}